# Network Slicing Management Technique for Local 5G Micro-Operator Deployments

Idris Badmus[1], Marja Matinmikko-Blue[1], and Jaspreet Singh Walia[2]
[1]Centre for Wireless Communication, University of Oulu, Oulu, Finland
[2]Department of Communications and Networking, Aalto University, Espoo, Finland

*Abstract*— Local 5G networks are expected to emerge to serve different vertical sectors' specific requirements. These networks can be deployed by traditional mobile network operators or entrant local operators. With a large number of verticals with different service requirements, while considering the network deployment cost in a single local area, it will not be economically feasible to deploy separate networks for each vertical. Thus, locally deployed 5G networks (aka micro operator networks) that can serve multiple verticals with multiple tenants in a location have gained increasing attention. Network slicing will enable a 5G micro-operator network to efficiently serve the multiple verticals and their tenants with different network requirements. This paper addresses how network slicing management functions can be used to implement, orchestrate and manage network slicing in different deployments of a local 5G micro-operator including the serving of closed, open and mixed customer groups. The paper proposes a descriptive technique by which different network slicing management functionalities defined by 3GPP can be used in coordination to create, orchestrate and manage network slicing for different deployment scenarios of a micro-operator. This is based on the network slice instance configuration type that can exist for each scenario. A network slice formation sequence is developed for the closed micro operator network to illustrate the tasks of the management functions. The results indicate that network slicing management plays a key role in designing local 5G networks that can serve different customer groups in the verticals.

## I. INTRODUCTION

The idea of the recently proposed 5G micro-operator concept is to deploy local cellular networks for vertical specific service delivery to provide enough network flexibility, customization and privacy required by the verticals [1]-[2]. In general, different verticals have varying and specific service requirements, which calls for local deployment of specialized networks. Network slicing, which is considered as a key enabler to a complete network softwarization in 5G [3], is the process of logically creating network slices, such that each slice will be responsible for different network requirements with adequate isolation and security [3]–[5]. To overcome the need for a large number of networks whereof each is tailored for the specifics of a certain vertical, network slicing allows the serving of the different verticals with different network requirements efficiently.

Several standards and commercial organizations have contributed towards defining network slicing. The NGMN alliance proposed a three layer model [5], and 3GPP defined the required network functions for network slice selection and management, and broadly defined three network slice types [6], [7]. Most research on network slicing [4], [8], [9] has been on regular Mobile Network Operator (MNO) networks with little emphasis on slicing in a locally deployed 5G network. The micro-operator network deployment scenarios can be closed, open and mixed [10]–[12]. According to [10], different NSI configuration types can exist for each deployment scenario. These NSI configuration types are used to describe how slicing in different deployment scenarios of a micro-operator network can be achieved. Meanwhile, to implement the network slicing for local 5G network, it is important to understand how different management functions can be used to achieve that. The Network slicing management functions such as the Network Slice Management Function (NSMF), Network Slice Subnet Management Function (NSSMF) and Communication Service Management Function (CSMF) are responsible for the end-to-end creation, management, and orchestration of network slice instance, network slice subnet instance, and managing the communication service and other network requirements in forming a slice. This is done while leveraging the use of SDN/NFV for logical slice creation [13].

This paper proposes a technique by which network slicing management functionalities can be used to implement network slicing in each deployment scenario of a local 5G micro-operator network based on the proposed Network Slice Instance (NSI) configuration types for the deployment scenarios. We present the generic network slicing management architecture for 5G micro-operator network and develop the slice formation sequence for the specific closed micro operator deployment.

The remaining part of this paper is arranged as follows. Section II discusses the previous work on local 5G micro-operator and the various deployment scenarios. Network slicing and network slicing management are discussed in Section III. Section IV proposes the architecture for the network slicing management technique and Section V presents the slice formation sequence for the closed micro-operator network. Finally, conclusions are drawn in Section VI.

## II. 5G MICRO-OPERATOR CONCEPT AND DEPLOYMENT SCENARIOS

Micro-operator concept aims at boosting vertical specific service delivery through locally deployed 5G network [2], [14]. The micro-operator can be deployed as a closed, open or mixed network [10]–[12], see Table I. A closed network, which can also be described as a vertical specific network [15], is targeted at verticals whose human and/or machine users present a closed users group. According to [10], a closed network can be of Deployment A where the micro-operator is serving one or more tenants with their operations within a

This work is supported by Business Finland in uO5G and MOSSAF project, and Academy of Finland in 6Genesis Flagship (grant no. 318927).

-single location or Deployment B where the micro-operator is serving one or more tenants whose operations are at multiple locations, and their communication will be enabled by connecting multiple tenants via an external network e.g. an MNO network. As defined in [10], the open network can serve MNOs' customers (MNO open network) or the general public. In the MNO open network case, the service agreement between the MNO and the micro-operator is such that the micro-operator will be responsible for network connectivity of the MNO subscribers within the locality. A single micro-operator can cover subscribers of multiple MNO using network slicing by dedicating different slices to different MNO subscribers. In the public open network, a separate slice is dedicated for general public subscribers within a locality, similar a public Wi-Fi network.

A mixed network is a micro-operator network which serves both closed and open groups of users with a defined level of privacy and isolation between the subscribers [10]. The mixed network can be deployed as Option A or Option B. Option A involves a network scenario where the micro-operator needs the MNO for specific wide area services such as accessing a faraway server, roaming, connecting closed tenants at different locations, etc. Option B refers to a scenario where the MNO needs the micro-operator network access to serve some of its subscribers within the micro-operator network coverage area. Each of these considered deployment scenarios can be realized with the help of network slicing, which in turn must be managed to achieve the requirements of the different user groups.

TABLE I. OVERVIEW OF DIFFERENT DEPLOYMENT SCENARIOS

| Micro-operator | Deployment Scenarios | Deployment Description |
|---|---|---|
| Closed | Deployment A | A closed network is responsible for one or more tenants at a single location. |
| Closed | Deployment B | A closed network is responsible for a set of tenants at different locations. |
| Open | MNO Open | Targeted at MNO subscribers within a given locality based on the service agreement with the MNO. The micro-operator may be responsible for subscribers from one MNO or from multiple MNO. |
| Open | Public Open | Targeted at the general public. |
| Mixed | Option A | Micro-operator needs services from the MNO such as wide area access, remote monitoring, etc. thereby configuring tenant's slices with MNO resources. |
| Mixed | Option B | MNO needs services from the micro-operator network such as better service for its subscribers within the micro-operator's network and using the tenant's broadband slice for extending indoor coverage. |

## III. NETWORK SLICE INSTANCES AND MANAGEMENT

According to 3GPP [7], an end-to-end NSI life cycle can be described in three stages. Starting from instantiation, configuration, and activation, followed by a run-time period for the slice which will involve supervision, reporting, and modification, and finally a slice deactivation and termination stage. For a micro-operator, the NSI configuration types [10] give an overview about how slice instances will be created for each deployment scenario. This will in-turn determine the type of services those deployment scenarios can cover. Three NSI configuration types proposed in [10] includes the Type 1, Type 2 and Type 3 NSI configuration. Type 1 NSI configuration involves having a network slice instance, such that there is no shared constituent between the Network Slice Subnet Instances (NSSI) or Network Functions (NF). The slice is created for tenants with strictly low latency and high reliability requirement or tenants that do not want to share network resources with other slices. The Type 2 NSI configuration involves slice creation for tenants with less strict latency requirement and thus can have shared NSSI or NFs between themselves. Type 3 NSI configuration is established for slices that need access to an external network, hence establishes a slice instance with NSSI from the external network e.g. a MNO network. The NSI configuration types for each of the deployment scenario can be seen in Fig. 1.

The resources to be used in achieving the NSI configuration types are virtualized and as such require proper management and orchestration to form a network slice [16]. In this light, 3GPP [7], described various management functions responsible for achieving network slicing. The CSMF is responsible for translating communication service requirement to network slice requirement for the NSMF based on the slice requests while providing and managing the communication service to the consumer. The NSMF is responsible for the management and orchestration of the NSI and communicating the network slice subnet requirements to the NSSMF. The NSSMF is responsible for the management and orchestration of the NSSI which contains the required NF. Thus, before a proper network slicing architectural implementation and design can be established for the various deployment scenarios of a micro-operator, it is vital to understand how the available management functions defined by 3GPP [7] can be designed or configured together to achieve such architecture.

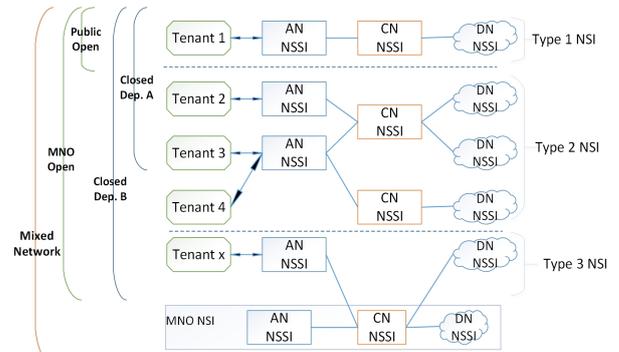

Fig. 1. NSI configuration for various deployment scenarios

## IV. PROPOSED NETWORK SLICING MANAGEMENT TECHNIQUE FOR MICRO-OPERATOR DEPLOYMENT SCENARIOS

To achieve a proper description of management technique for network slicing, we propose a generic management architecture (Fig. 2) for the end-to-end implementation of network slicing management function for the micro-operator network while considering the various deployment scenario. In the proposed management technique, resource selection for various vertical slices can happen within a single domain or multiple domains [17] and can be managed and controlled by single or multi-domain administrator [18]. Also, the proposed architecture in Fig. 2 contains implementation description for all the micro-operator deployment scenarios listed in Table I combined together, and as such, it can easily be rearranged or renamed for any individual deployment scenario with proper description.

The architecural blocks are classified into three categories, namely the basic blocks, managemement blocks and the resource blocks. The basic contains the tenants, network provider and the service provider while the managemement blocks which entails all the management functionalities such as the CSMF, the NSMF and the NSSMF, and finally the resource blocks, which contain the NF resources, and the configured NSSI and NSI. Indiviaul blocks are described in more details below.

*A. Basic blocks*

**Tenants block** will serve as the communication service consumer. The tenant's User Equipment (UE) send the slice request based on the communications service requirements. Every slice request made by a tenant contains the tenant slice ID, tenant request message (network requirement), tenant sharing agreement (shared constituent or not), location of other tenants to share constituents with (based on deployment scenario, e.g. Dep. A if shared constituent is in same location, Dep. B if shared constituent is in different location. Tenant embeds all information into the slice request and sends slice request to both the comm. service provider and CSMF.

**Communication service provider** can be a separate entity like a slice broker [19] or part of the micro-operator's network. It gets slice request from all different tenants and forwards slice requirements to the network provider. It manages the communications services using CSMF and provides the requested service to the tenant.

**Micro-operator** serves as the network provider, and it manages both the NSMF and NSSMF. It provides approval for slice creation to the NSMF and the NSSMF based on the slice request ID received from the comm. service provider. Generally, when a slice request comes from the comm. service provider to the network provider and service agreement with the tenant is valid, the micro-operator will approve the slice request and the formation of NSSI and NSI, based on tenant policy, charging, subscription and sharing info. The micro-operator then allocates resources to the NSSMF and NSMF based on the network requirements such as, latency, throughput, network sharing, mobility, location and slice duration for the creation of NSSI and NSI.

*B. Management blocks*

**CSMF** manages and controls the activities of both the tenants and comm. service provider by converting communication requirement to network slice requirement for all deployment scenarios. The CSMF connects to the NSMF to get the NSI as a communication service and based on the NSIs provided by the NSMF, the CSMF transmit each NSI as a communication service to the exact tenant. Furthermore, In case of mixed network option B where the MNO needs the micro-operator (i.e. tenant slice will be composed of both MNO NSI and micro-operator NSI), the CSMF gets NSI from both the micro-operator NSMF and MNO NSMF and manage both NSIs as the comm. Service which is to be allocated to the tenant. The CSMF also gets NSI activation and termination from the NSMF.

**NSMF** is one of the most important network slicing management functions. The NSMF is responsible for the co-ordination, management, and orchestration of NSI (i.e. end-to-end NSI life cycle management). The NSMF gets a slice request from the CSMF and wait for slice creation approval from the network provider and if there is no approval it discards the slice request. After getting approval, the NSMF will instantiate the NSI formation based on the embedded slice request information. For Closed network, it checks if the NSI will contain a shared NSSI constituent with MNO network (i.e. Dep. B) for wide area access or roaming. If it will, it connects with the MNO NSSMF to get the required NSSI and gets the remaining NSSI from the micro-operator NSSMF before forming the NSI. However, if the slice request does not involve an external network, then all shared or non-shared constituent will be obtained from the micro-operator's NSSMF. This will also depend on the locations of the tenants whose slices will share resources. If these tenants are all within the same location (i.e. Dep. A), then all NSSI involved in making the NSI will be obtained from the micro-operator NSSMF at that location, otherwise it's Dep. B where some NSSI will be obtained from other locations. For Open network, the NSMF will make NSI available for MNO based on the service agreement with the micro-operator. Unlike the closed network, in the Open network, the distinction from an MNO open network or a public open network will be determined from the comm. service provider, and the micro-operator will just give the approval for a slice creation. Also, the network provider can also connect with the MNO to confirm subscriber policy. In the mixed network, all the previous cases will exist. Also, if the MNO needs access the micro-operator's network, then the CSMF will make the tenant slice from both MNO and micro-operator's NSI. However, if the micro-operator needs access to the MNO's network, the NSMF will make the NSI from both MNO and micro-operator's NSSI.

**NSSMF** is responsible for the management and orchestration of the NSSI throughout its lifecycle. It also gets request approval from the network provider after an NSSI request from the NSMF. Based on the per-slice request, the NSSMF select the required resources in forming either the AN, CN or DN NSSIs. NSSMF will get all the required resources in making the NSSI from the Network functions (NFs) then manage and control the transmission of individual NSSIs to form an NSI. Deactivated resources can be dynamically removed and replaced, or reallocated to another NSSI by the NSSMF. In the selection of network functions, the NSSMF can select network resources from a single or multiple network domains and aggregate them together as described by [18].

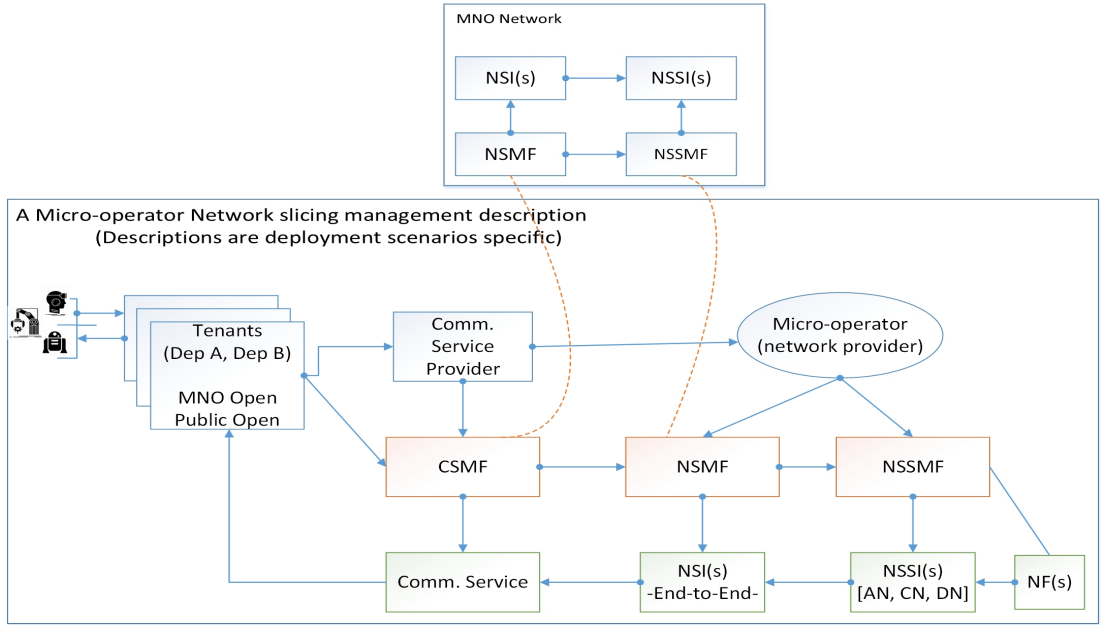

Fig. 2. Network slice management technique for a micro-operator's network

## C. Resource Blocks.

**NF** will contain Virtual Network Functions (VNFs) and Physical Network Functions (PNFs) resources whose managements are described by ETSI [20]. NFs selected will form the NSSI requested by the NSSMF and it might also be reallocated to other NSSI after the slice is terminated.

**NSSI** is the output from the NSSMF, and all NSSIs are transmitted to the NSMF in forming the NSI. The **NSI** will entail the end-to end slice instances for multiple tenants. It is managed and orchestrated by the NSMF and it is transmitted as a communication service to the tenants. The NSI transmitted will be specific for each tenant.

The NSI attributed to each tenant is described as the **communication service**. It is managed by the CSMF, which will select the communication service that should be enabled for individual tenant. The comm. service contains end-to-end network connectivity per tenant throughout the slice creation. Finally, a network slice can be attributed to a tenant.

## V. NETWORK SLICE FORMATION SEQUENCE FOR CLOSED MICRO OPERATOR NETWORK

From the management description of a micro-operator network in Fig. 2, it can be seen that it will be important to create a slice formation sequence for the different deployment scenarios. We define a slice formation sequence as a sequence of operations for slice creation from the tenant's slice request to the tenant slice formation. Basically, creating a slice formation sequence will give a better view on how network slicing management coordinates slice creation for each deployment. The slice formation sequence will be handled separately for each deployment scenarios by the network slicing management functions, and this will make it easy to have a laid down management description per deployment scenario, easing the slice formation process. Thus, only the resource selection will be different and that is based on the required service by the tenant. It will also reduce extensively the slice creation period for the micro-operator network compared to the MNO network whose slice formation sequences are to be formed for every new slice.

With each block previously explained in Fig. 2, Fig. 3 describes in detail the slice formation sequence for a closed micro-operator network. The formation sequence steps is represented with numbers from 0-15. Step 0 is from the UEs to the tenants when the UEs are waiting for connection to be enabled by the tenant. The main network slice formation starts from Step 1 when the closed tenants with the tenant slice IDs send their slice request information to the comm. Service provider and the CSMF. Step 2 describes when the comm. service provider, operating the CSMF, transmits per tenant slice request to the network provider. Based on the tenant slice request transmitted to the network provider, the CSMF in Step 3, transmit per tenant request info to the NSMF for NSI creation. Network provider which contains both NSMF and NSSMF in Step 4 approves the slice creation and provides the network requirement for the tenant to the NSMF and NSSMF. If, based on the slice request and service requirement received by NSMF, the slice falls into Dep. A, NSMF in Step 5 requests NSSI from the micro-operator NSSMF, but if it belongs to Dep. B, NSMF request for NSSI from both micro-operator (Step 5) and MNO NSSMF (Step 6). Based on service requirement and slice requirement Step 7 describes when NSSMF request all needed VNF/PNF resources from the NFs (managed by a VNF orchestrator defined by ETSI [20]). Step 8 describe resources being provided as NSSIs, all managed by the NSSMF for shared or non-shared constituents (Step 9).

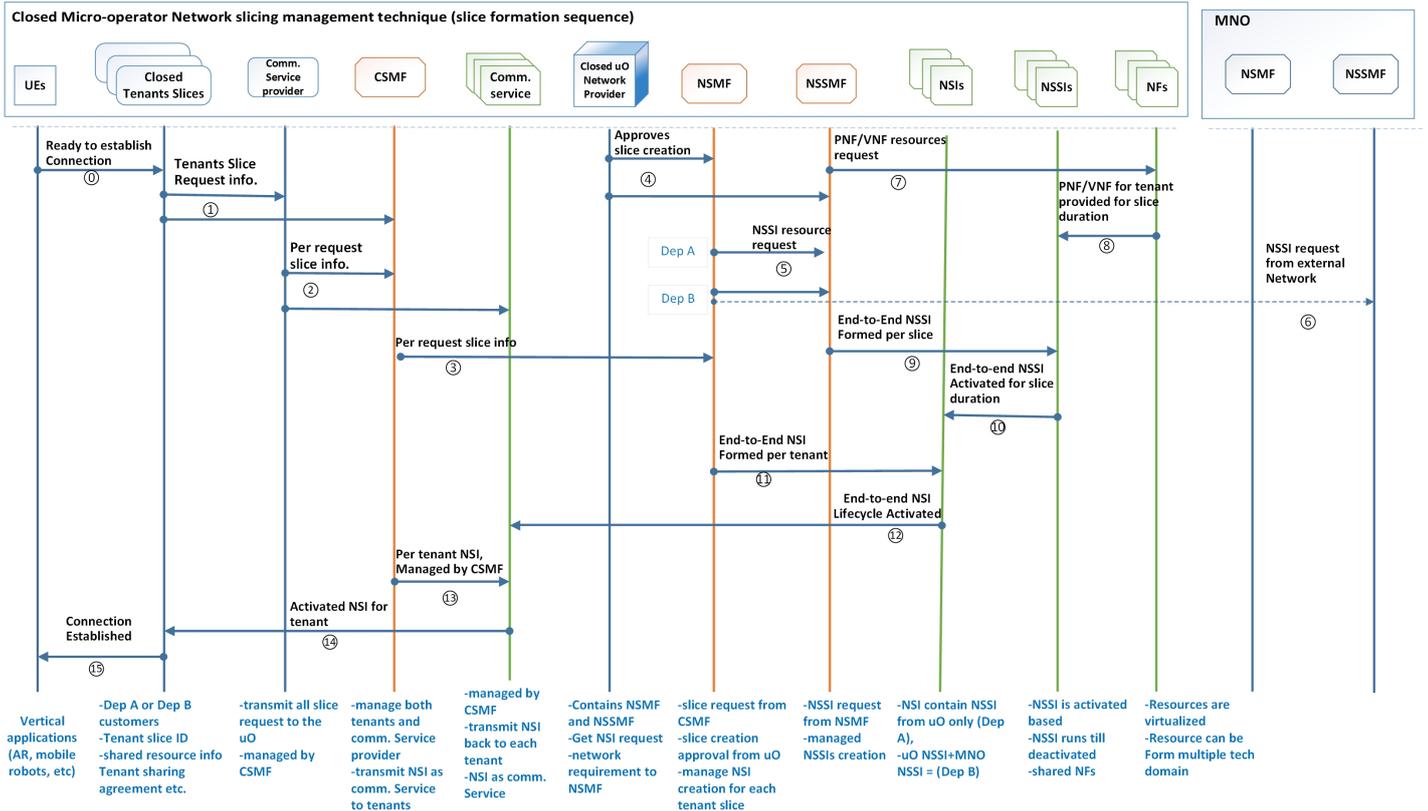

Fig. 3. Slice formation sequence for a closed micro-operator's network

NSSIs are provided as NSI in Step 10, managed by NSMF to be attributed back to individual tenant (Step 11). Activated end-to-end NSI for each tenant, based on the deployment types, is transmitted as communication services in Step 12, now managed by CSMF for each tenant (Step 13). Step 14 shows communication service provided to individual tenant and Step 15 describes the connection established by the tenants for the UEs.

## VI. CONCLUSION

A proper description of network slicing for a local 5G micro-operator cannot be over emphasized as it will ease its deployment and will make it economically feasible and easy for any stakeholder to deploy a private network that can cover verticals of different network requirements. This paper has presented how the various network slice management functions defined by 3GPP can be used to achieve the network slicing for the various deployment scenarios of a micro-operator. This paper has shown how the management functions can be architecturally designed to achieve slicing for the various deployment scenarios of a micro-operator and formed the required slice formation sequence.